\newcommand{\ud}{\,\mathrm{d}}

\documentclass{article}

\usepackage{amsmath}
\usepackage{graphicx}
\usepackage{mathrsfs}

\title{Snapshot projection optical tomography}

\author{
  Yongjin Sung\\
  College of Engineering \& Applied Science\\
  University of Wisconsin\\
  Milwaukee, WI 53211\\
  \texttt{ysung4@uwm.edu}\\
}

\begin{document}
\maketitle

\begin{abstract}
We present a plenoptic microscopy configuration for 3D snapshot imaging, which is dual telecentric and can directly record true projection images corresponding with different viewing angles. It also allows blocking high-angle stray rays without sacrificing the light collection efficiency. This configuration named as snapshot projection optical tomography (SPOT) arranges an objective lens and a microlens array (MLA) in a 4-f telecentric configuration and places an aperture stop at the back focal plane of a relay lens. We develop a forward imaging model for SPOT, which can also be applied to existing light field microscopy techniques using an MLA as tube lens. Using the developed system, we demonstrate snapshot 3D imaging of various fluorescent beads and a biological cell, which confirms the capability of SPOT for imaging specimens with an extended fluorophore distribution as well as isolated fluorochromes. The transverse and vertical resolutions are measured to be 0.8 $\mu$m and 1.6 $\mu$m, respectively. 
\end{abstract}

\section{Introduction}
Fluorescence microscopy allows for imaging the distribution of biomolecules within a cell, tissue, or animal, and it has become an essential tool in biomedical researches and clinical practices\cite{masters_principles_2008}. A variety of new molecular probes and gene manipulation techniques are adding more functionalities to fluorescence microscopy. To record the 3D distribution of fluorophores, whether they are reactive dyes, quantum dots, or fluorescent proteins, three strategies are typically used. In the first approach, a focused beam\cite{sheppard_confocal_1997} or a light sheet\cite{keller_reconstruction_2008} scans the sample volume, and, for each beam position, only the fluorophores within the excitation volume contribute to the image. In the second approach, the entire sample volume is illuminated with excitation light, while the objective lens with a shallow depth of field creates a sharp image only for a certain depth at a time\cite{sarder_deconvolution_2006}. In the third approach, the volumetric excitation is combined with an imaging system with a large depth of field, so that the recorded image can be related to a projection of the 3D fluorophore distribution along the optical axis. The so-called projection images are acquired for varying orientations of the sample, then combined to provide the 3D fluorophore distribution in the sample volume\cite{sharpe_optical_2002} as in X-ray computed tomography (CT)\cite{kak_principles_1988}. This method is called optical projection tomography\cite{sharpe_optical_2002}. Whichever method is used, for 3D imaging, a series of data is acquired, which can take a fraction of a second even with a fast camera recently developed. Because samples can move or change their shape during the data acquisition, these scanning-based methods are not ideal for high-speed 3D imaging. 

Snapshot volumetric imaging techniques have been developed to overcome this limitation\cite{gao_review_2016}. Typically, less than 10 depth layers are remapped to different areas of the camera using a volume hologram\cite{sinha_imaging_2004}, a distorted grating\cite{blanchard_simultaneous_1999}, or a liquid-crystal spatial light modulator\cite{maurer_depth_2010}. Simultaneous imaging of 49 layers at a coarse axial resolution of 38 $\mu$m has also been demonstrated using a distorted Dammann grating\cite{yu_distorted_2013}. Most notably, light-field microscopy (LFM)\cite{levoy_light_2006} uses a microlens array (MLA) to record a 4D light field (i.e., 2D spatial and 2D angular distribution of intensity among light rays), from which the projection images required for tomographic reconstruction are synthesized. Originally derived from plenoptic camera\cite{adelson_single_1992}, LFM has opened a way to high-speed volumetric imaging and has been adopted to record neuronal activities of \textit{C. elegans} and small zebrafish\cite{prevedel_simultaneous_2014,pegard_compressive_2016}. Of note, the spatial resolution of LFM is not sufficient for cell imaging, and the transverse resolution is known to vary across the depth\cite{levoy_recording_2009}. Wavefront coding, i.e., engineering the point spread function using a phase mask, has been shown to mitigate this problem\cite{cohen_enhancing_2014}.
Recently, the second class of LFM promising a better resolution characteristic has been proposed, which places an MLA at the Fourier plane instead of the image plane. Examples are Fourier integral microscopy (FiMic)\cite{llavador_resolution_2016,scrofani_fimic:_2018}, Fourier light-field microscopy (FLFM)\cite{guo_fourier_2019}, and eXtended field of view LFM (XLFM)\cite{cong_rapid_2017}. FiMic and FLFM use the same optical configuration, although the image reconstruction is different. XLFM places an aperture array in front of MLA, which renders the system to be object-side telecentric at the cost of reduced light collection efficiency. In this paper, we present a plenoptic configuration, which may be classified as the third class of LFM. This configuration named as snapshot projection optical tomography (SPOT) places an MLA in a 4-f telecentric configuration with the objective lens and installs a single aperture stop at the back focal plane of a relay system, which is also in a 4-f configuration. SPOT is dual telecentric (i.e., both object- and image-side telecentric) and records true projection images, thereby allowing for a fast and an accurate tomographic reconstruction.
\begin{figure}[htbp]
\centering
\includegraphics[width=\linewidth]{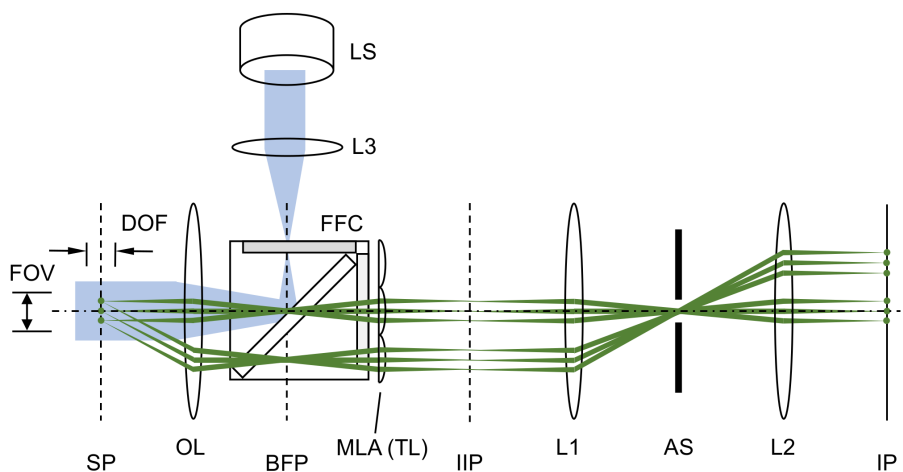}
\caption{Snapshot projection optical tomography (SPOT). The excitation beam path from the light source is shown in blue, and the example emission paths from three fluorophores in the sample plane are shown in green. SP: sample plane; BFP: back focal plane of the objective lens; IIP: intermediate image plane; IP: image plane; OL: objective lens; MLA: microlens array; TL: tube lens; AS: aperture stop; LS: light source; L1, L2, and L3: lenses; FFC: fluorescence filter cube; FOV: field of view; and DOF: depth of field. The dash-dot line represents the optical axis of objective lens.}
\label{fig1}
\end{figure}

Figure~\ref{fig1} shows a schematic diagram of SPOT. In the figure, the emission beam paths through the on-axis lenslet and those through one of the off-axis lenslets are shown together. The depth of field of an optical microscope is inversely proportional to the square of the numerical aperture (NA) of objective lens. When a 2D array of lenslets is used to capture $m \times m$ projection images, the effective NA for each projection image is reduced to NA/$m$. Thereby, the depth of field is significantly increased, and the image recorded by each lenslet can be considered as the accumulation of the fluorescence signal along a certain viewing direction. As is shown in Fig.~\ref{fig1}, the projection images are formed at the intermediate image plane (IIP), which are relayed to the image plane (IP) by the lenses L1 and L2. The iris diaphragm placed at the back focal plane of L1 serves as the aperture stop for each pair of the objective lens and one of the lenslets; it controls the amount of light passing through each lenslet. The aperture stop is placed at a plane conjugate to the focal planes of objective lens and MLA (tube lens), which renders SPOT to be dual telecentric. As will be shown in the Discussion section, it also blocks high-angle rays emitted from the fluorophores outside the field of view, which can generate a ghost image otherwise. This capability is especially imporatnt for imaging a sample with high population density. 

\section{Theory}
\begin{figure}[htbp]
\centering
\includegraphics[width=\linewidth]{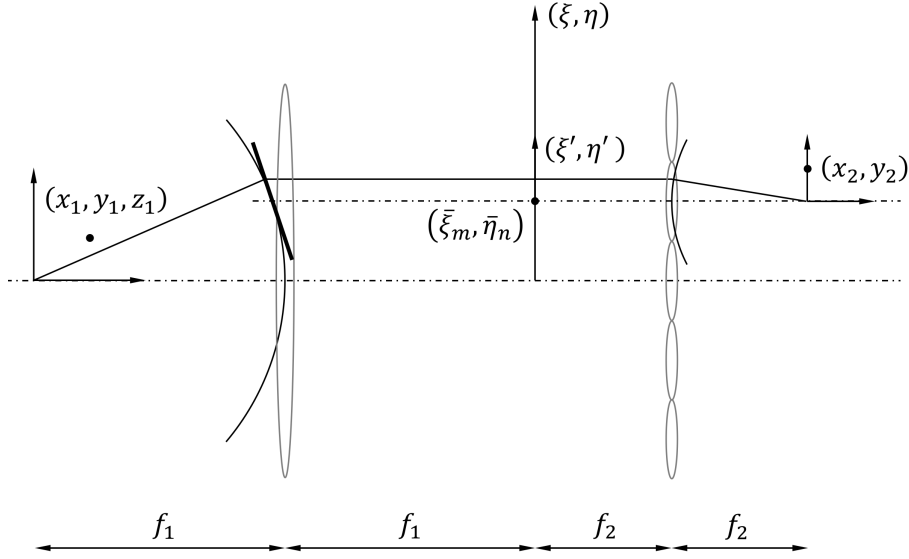}
\caption{Schematic diagram of the imaging geometry consisting of a high-NA objective lens and a 2D array of lenslets.}
\label{fig2}
\end{figure}

In this section, we develop an image formation model for SPOT. Figure~\ref{fig2} shows a schematic diagram of SPOT, which consists of a high-NA objective lens and an MLA in a 4-f telecentric configuration. The forward imaging model consists of two parts: one describing the light collection by the objective lens, and the other describing the image formation by the lenslets. High NA is typically related to high transverse resolution. In SPOT, it is also important for high depth resolution, because NA determines the maximum viewing angle of projection image. Thus, for the first part of the forward imaging model, we adopt a non-paraxial theory\cite{sheppard_imaging_1993}, which can be applied to a high-NA objective lens. The rays captured by the objective lens are relayed to the image plane by the lenslet array. Noteworthy, the rays through each lenslet subtend only a small solid angle at the focused point in the image plane; the effective NA for individual projection image is much smaller than the NA of objective lens. Thus, we can introduce a small-angle approximation in the second part of the forward imaging model, which helps to simplify the formulation with minimally sacrificing the accuracy. 

Suppose that a point source of amplitude $U_1$ is placed at $(x_1,y_1,z_1)$ in the sample space and radiates isotropically. The amplitude in the pupil plane with the coordinates $\xi$ and $\eta$ can be written as
\begin{equation}
U(\xi,\eta) = -\frac{i a_1^{-1}(\xi,\eta)}{\lambda f_1} U_1 \text{exp} \left\{ \frac{i k}{f_1} \left (x_1 \xi + y_1 \eta + z_1 \zeta \right) \right\}, 
\end{equation}
where $f_1$ and $f_2$ are the focal lengths of the objective lens and MLA, respectively, and $\zeta = [f_1^2-(\xi^2+\eta^2)]^{1/2}$. The apodization factor $a_1(\xi,\eta)$ is $\left[ 1-(\xi^2+\eta^2)/f_1^2 \right]^{1/4}$ for the sine condition and 1 for the Herschel condition. Equation (1) is valid for large $\xi$ and $\eta$ values.

Next, consider the image formed by the $(m,n)$th lenslet, whose center is located at $(\bar{\xi}_m,\bar{\eta}_n)$. We can define a local coordinate system $(\xi',\eta')$ in the pupil plane with the origin at the lenslet center, i.e., $(\xi',\eta')=(\xi-\bar{\xi}_m,\eta-\bar{\eta}_n)$. Note that the rays captured by the lenslet and contributing to the image subtend only a small angle at the image plane, which can justify $\xi'\ll1$ and $\eta'\ll1$. Then, the variable $\zeta$ in Eq. (1) can be approximated as $\zeta \approx c \left ( f_1^2- \bar{\xi}_m \xi - \bar{\eta}_n \eta \right )$, where $c = \left [f_1^2-({\bar{\xi}_m}^2+{\bar{\eta}_n}^2) \right ]^{-1/2}$. A geometrical interpretation of this simplification is to approximate a spherical surface with a plane that is tangent to the sphere at the point $(\bar{\xi}_m,\bar{\eta}_n,[f_1^2-({\bar{\xi}_m}^2+{\bar{\eta}_n}^2)]^{1/2})$, as shown in Fig.~\ref{fig2}. Again, this approximation is valid as far as the NA of lenslets is small, even though the NA of objective lens can be very large. Then, in the coordinates $(\xi',\eta')$, Eq. (1) can be written as
\begin{multline}
U'(\xi',\eta') = -\frac{i a_1^{-1}(\xi'+\bar{\xi}_m,\eta'+\bar{\eta}_n)}{\lambda f_1} U_1 \text{exp} (i k f_1 z_1 c) \\
\times \text{exp} \left\{ \frac{i k}{f_1} \left [ (x_1 - \Delta x_1) (\xi'+\bar{\xi}_m) + (y_1 - \Delta y_1) (\eta'+\bar{\eta}_n) \right ] \right\}, 
\end{multline}
where $\Delta x_1=z_1 \left\{ {\bar{\xi}_m}/{\left [f_1^2-({\bar{\xi}_m}^2+{\bar{\eta}_n}^2) \right ]^{1/2}} \right\}$ and $\Delta y_1=z_1 \left\{ {\bar{\eta}_n}/{\left [f_1^2-({\bar{\xi}_m}^2+{\bar{\eta}_n}^2) \right ]^{1/2}} \right\}$. 

For the amplitude distribution $U'(\xi',\eta')$ in the pupil plane, the lenslet centered at $(\bar{\xi}_m,\bar{\eta}_n)$ forms an image at the image plane, which can be written as
\begin{multline}
U_2(x_2,y_2) = -\frac{i}{\lambda f_2} \iint U'(\xi',\eta') P(\xi',\eta') a_2(\xi',\eta') \\
\times \text{exp} \left\{ \frac{i k}{f_2} \left (x_2 \xi' + y_2 \eta' \right) \right\} \ud \xi' \ud \eta',
\end{multline}
where $P(\xi',\eta')$ is the pupil function, which includes optical aberration. The apodization factor $a_2(\xi',\eta')$ is $\left[ 1-({\xi'}^2+{\eta'}^2)/f_2^2 \right]^{1/4}$ for the sine condition and 1 for the Herschel condition.
Substituting Eq. (2) and excluding a phase factor, Eq. (3) can be written as
\begin{multline}
U_2(x_2,y_2) = -\frac{U_1}{\lambda^2 f_1 f_2} \\
\times \Psi \left(- \frac{x_2+M(x_1-\Delta x_1)}{\lambda f_2} ,- \frac{y_2+M(y_1-\Delta y_1)}{\lambda f_2} \right),
\end{multline}
where $M=f_2/f_1$ is the magnification, and
\begin{multline}
\Psi(u,v)=\mathscr{F} \left\{ \frac{a_2(\xi',\eta')}{a_1(\xi'+\bar{\xi}_m,\eta'+\bar{\eta}_n)} P(\xi',\eta') \right\} \\
=\iint\displaylimits_{-\infty}^{\infty} \frac{a_2(\xi',\eta') P(\xi',\eta')}{a_1(\xi'+\bar{\xi}_m,\eta'+\bar{\eta}_n)} \text{exp} \{-i2\pi(u\xi'+v\xi') \} \ud \xi' \ud \eta'.
\end{multline}

Using the sine condition, $\Psi(u,v)$ is simply the 2D Fourier transform of the pupil function $P(\xi',\eta')$. The aberration-free pupil function for a rectangular aperture can be expressed as $P(\xi',\eta')=rect(\xi'/p)rect(\eta'/p)$, where $p$ is the width of lenslet, and $\text{rect}(x)=1$ (for $\lvert x \rvert < 1/2$); $1/2$ (for $\lvert x \rvert = 1/2$); and $0$ otherwise. The 2D Fourier transform of $P(\xi',\eta')$ is $\tilde{P}(u,v)=p^2 \text{sinc}(pu)\text{sinc}(pv)$, where $\text{sinc}(x)=\text{sin}(\pi x)/(\pi x)$. Then, the intensity at the image plane can be expressed as
\begin{multline}
I_2(x_2,y_2) \propto \text{sinc}^2 \left(\frac{x_2+M(x_1-\Delta x_1)}{\lambda f_2 / p} \right) \\
\times \text{sinc}^2 \left(\frac{y_2+M(y_1-\Delta y_1)}{\lambda f_2 / p} \right).
\end{multline}

Equation (6) is the point spread function for an imaging system consisting of an objective lens and an off-axis lenslet. Setting aside the magnification and the inversion, we can consider the image formation as a two-step process: shifting of the fluorophore by $\Delta x_1$ and $\Delta y_1$ along the $x$ and $y$ directions, respectively, and blurring due to the small aperture of the lenslet. Note that $\Delta x_1$ and $\Delta y_1$ can be written as $\Delta x_1=z_1 \text{tan} (\alpha_m)$ and $\Delta y_1=z_1 \text{tan} (\beta_n)$, where $\alpha_m$ and $\beta_n$ are given by
\begin{subequations}
\begin{eqnarray}
\alpha_m = \text{tan}^{-1} \left\{ \frac{\bar{\xi}_m}{\left [f_1^2-({\bar{\xi}_m}^2+{\bar{\eta}_n}^2) \right ]^{1/2}} \right\},\\
\beta_n = \text{tan}^{-1} \left\{ \frac{\bar{\eta}_n}{\left [f_1^2-({\bar{\xi}_m}^2+{\bar{\eta}_n}^2) \right ]^{1/2}} \right\}.
\end{eqnarray}
\end{subequations}

\begin{figure}[htbp]
\centering
\includegraphics[width=\linewidth]{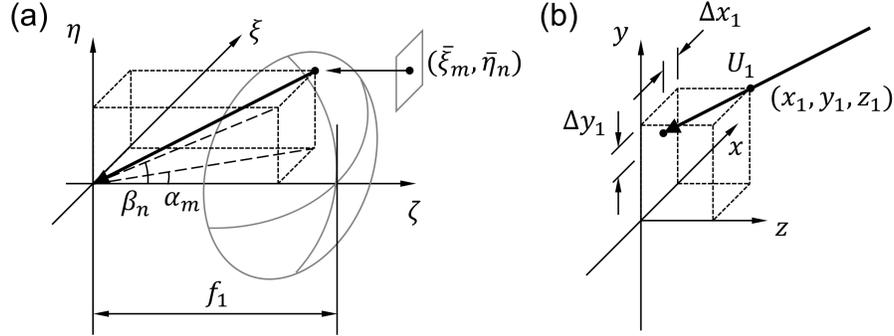}
\caption{Geometrical interpretation of the projection operation performed by an off-axis lenslet in SPOT. (a) The projection direction is determined by the offset of lenslet center and the focal length of objective lens. (b) An out-of-focus point is projected onto the $z=0$ plane along the projection direction determined in (a).}
\label{fig3}
\end{figure}

In Fig.~\ref{fig3}(a), $\alpha_m$ and $\beta_n$ are the angles that the projections of the arrow onto the $\xi-\zeta$ and $\eta-\zeta$ planes, respectively, make with the $\zeta$ axis. These angles are determined by the focal length of the objective lens and the offset of the lenslet center from the optical axis. Thus, the direction of arrow may be considered as the viewing direction of the lenslet. In Fig.~\ref{fig3}(b), a point at $(x_1,y_1,z_1)$ is projected onto the $x-y$ plane along the direction of arrow shown in Fig.~\ref{fig3}(a). $\Delta x_1$ and $\Delta y_1$ are the transverse shifts of the point along the $x$ and $y$ directions, respectively. Thus, we may consider the shifting operation in Eq. (6) as the projection operation along the lenslet's viewing direction. The blurring operation can be described by a convolution with a Gaussian filter, whose full-width-at-half-maximum (FWHM) is $0.443 \lambda / \text{NA}_\text{MLA} / M$, where $M$ is the magnification, and $\text{NA}_\text{MLA}$ is the NA of the MLA. Using the projection and blurring operators, Eq. (6), the image of a point-like fluorophore recorded by the $(m,n)$th lenslet, can be rewritten as
\begin{equation}
I^{(m,n)}(x,y) = \left( P_{m,n} \delta (x-x_i,y-y_i,z-z_i) \right) * h(x,y),
\end{equation}
where $P_{m,n}$ and $h(x,y)$ represent the projection and blurring operators, respectively. For a general specimen with extended fluorophore distribution, the projection image can be written as
\begin{equation}
I^{(m,n)}(x,y) = \left( P_{m,n} O(x,y,z) \right) * h(x,y).
\end{equation}

The image formation model developed here can also be applied to FiMic/FLFM, which place an MLA at the back focal plane of objective lens. It is because the change in the propagation distance between the objective lens and tube lens does not alter the light intensity distribution, as long as the camera is located at the back focal plane of MLA\cite{goodman_introduction_2004}. In contrast, XLFM places two groups of lenslets at different distances from the pupil plane, and thus the camera is not at the back focal plane of either lenslet group\cite{cong_rapid_2017}. Thus, the images recorded with XLFM cannot be represented by Eq. (9).

\section{Methods}
We built a prototype SPOT system on an optical breadboard. A fluorescence illuminator equipped with a triple bandpass filter (Chroma, 69002) was obtained from a cannibalized microscope (Zeiss, Axiostar) and incorporated into the prototype system. The excitation lamp was replaced with a high-power (3.3 W), broadband (450-670 nm) light-emitting diode (Thorlabs, SOLIS-1C). We used a high-NA objective lens (Olympus, UPlanFl 100X, 1.3 NA) and an MLA (Edmund, 86-745) as a tube lens, which had the focal length of 4.8 mm and the pitch of 300 $\mu$m. The back focal plane of objective lens was inside the barrel and the focal length of MLA was not long enough; thus, to arrange the objective lens and MLA in a 4-f configuration, we inserted two (4-f) relay lenses between them, which is not shown in Fig.~\ref{fig1}. The excitation beam was introduced into the sample plane through the filter cube installed at a Fourier plane (i.e., a plane conjugate to the back focal plane) of the objective lens. The fluorescence light emitted from the sample was captured by the objective lens, and an image was formed at the back focal plane of MLA. After the MLA, we installed two relay lenses of focal lengths 85 mm and 105 mm in a 4-f configuration, and an iris diaphragm at the back focal plane of the first relay lens. A charge-coupled device camera (Allied Vision, Stingray F-504C) installed at the image plane (i.e., the back focal plane of the second relay lens) had $2452\times2056$ pixels and 3.45 $\mu$m pitch. The overall magnification was 6.7, the field of view was 56 $\mu$m, and the camera pixel resolution was 0.5 $\mu$m. For these choices of parameters, the effective NA for the projection images was 0.21. The diameter of the aperture stop was adjusted to about 5 mm, which corresponded with the maximum size of the ray bundles at the location. For the center wavelength (505 nm) of GFP emission band, the diffraction limit is 1.2 $\mu$m. The pixel resolution was higher than the Nyquist sample rate for the diffraction limit. To record a raw SPOT image, we used the maximum gain (670) and variable exposure time to increase the signal intensity while preventing sensor saturation. For bright-field and wide-field fluorescence imaging, a separate beam path was installed in tandem with the SPOT system. Switching between the imaging modes was facilitated using a flip mirror installed in the shared beam path. To demonstrate 3D imaging using the developed system, we first imaged a 20 $\mu$m fluorescent bead (Polysciences, 19096-2) with yellow-green fluorescent dye incorporated into the center core. The actual size of the fluorescent core was variable and different from the nominal value of the bead's diameter. Next, we imaged 6-$\mu$m polystyrene beads (Thermo Fisher, F14806), whose outermost layers were stained with green/orange/dark-red fluorescent dye. For the measurement of 3D PSF, we used 0.5 $\mu$m fluorescent beads (Thermo Fisher, F13839). For all the bead measurements, we added to the sample immersion oil (Olympus, IMMOIL-F30CC) with the refractive index of 1.518 as surrounding medium.
\begin{figure}[htbp]
\centering
\includegraphics[width=\linewidth]{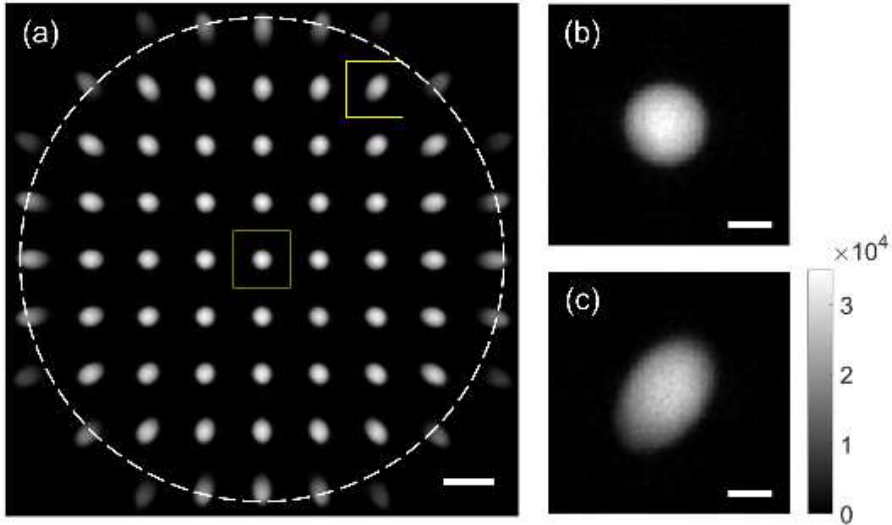}
\caption{Example raw image (a) acquired with SPOT.Two images in the rectangles are magnified on the right (b and c). Scale bars: 50 $\mu$m in (a) and 10 $\mu$m in (b) and (c).}
\label{fig4}
\end{figure}

Figure~\ref{fig4}(a) shows an example raw image acquired with SPOT. It consists of a multitude of projection images corresponding with different viewing angles. Two of the projection images in the dotted boxes are magnified on the right (b and c). The sample was a 20 $\mu$m fluorescent bead. The dotted circle in Fig.~\ref{fig4}(a) represents the NA of the objective lens. The projection images on the circle showed vignetting and were discarded. The projection images are elongated along different viewing directions, as shown in Fig.~\ref{fig4}(c). The 45 projection images inside the NA were extracted and saved separately for further processing. The size and location of the window used for the extraction can be determined based on visual inspection of the raw image, which may cause misalignment between the extracted images. To mitigate this problem, we applied intensity-based image registration to each projection image.

\begin{figure}[htbp]
\centering
\includegraphics[width=\linewidth]{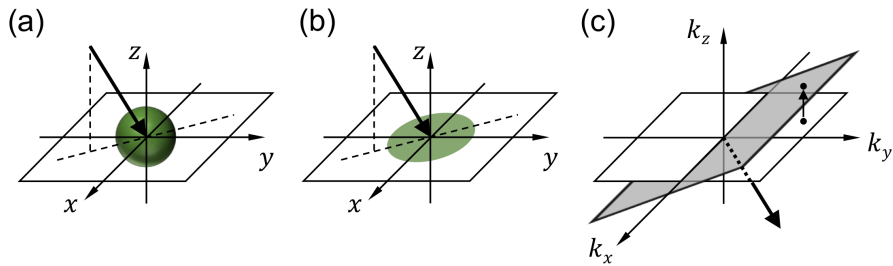}
\caption{Projection operation performed by an off-axis lenslet and inverse projection of the image: (a) imaging geometry; (b) projection operation; and (c) inverse projection operation.}
\label{fig5}
\end{figure}

The theory developed in the previous section describes the image formation by each lenslet as a two-step process: projection of the 3D fluorophore distribution onto the sample plane and convolution with a blurring function. Similarly, the 3D fluorophore distribution can be reconstructed from the recorded projection images in a two-step process: deconvolution and inverse projection. For the deconvolution, we used the Richardson-Lucy method\cite{lucy_iterative_1974,richardson_bayesian-based_1972}, which was implemented as a built-in function in Matlab (Mathworks, 2018a). The inverse projection operation is typically done using the inverse Radon transform, which is also illustrated in Fig.~\ref{fig5} and explained below in more detail. 

According to the Fourier slice theorem\cite{kak_principles_1988}, the projection of an object (i.e., the accumulation of the pixel values within the object along a certain direction) in the physical space can be related to the object's spatial frequency spectrum in the Fourier space. The Fourier slice theorem is the basis of the filtered back-projection method, which is one of the fastest methods for tomographic reconstruction\cite{kak_principles_1988}. It can be combined with an iterative reconstruction method to provide a fast yet accurate result\cite{beister_iterative_2012}. These approaches originally developed for X-ray CT are also valid in optical imaging when the projection images are recorded using a low (effective) NA optical system\cite{sharpe_optical_2002}. Noteworthy, the original Fourier slice theorem assumes that the projection direction is perpendicular to the plane onto which the object is projected. As shown in Fig.~\ref{fig5}, the images recorded by SPOT are the projections of the object onto the sample plane, which is not perpendicular to the projection direction. Therefore, to apply the Fourier slice theorem, the 2D Fourier transform of each projection image, which provides the spatial-frequency components on the $k_x-k_y$ plane, needs be projected vertically onto the tilted plane, as is shown in Fig.~\ref{fig5}(c). The surface normal vector of the tilted plane is the lenslet's viewing direction. Thus, the other projection images corresponding with different viewing angles are projected onto the planes with different orientations. After completing the mapping in the 3D spatial-frequency space, we can simply take the 3D inverse Fourier transform to provide the 3D fluorophore distribution within the object.

Noteworthy, wide-field fluorescence imaging, which includes LFM, FiMic/FLFM, XLFM, and SPOT, suffers from the missing-cone artefact as the maximum viewing angle is limited by the finite NA of objective lens. This contrasts with CT, in which a pair of X-ray source and detector is rotated around a patient, so that the projection images for all viewing directions can be collected. We note that CT can still have a missing-angle artefact, which usually refers to the missing data due to uniform undersampling in the Fourier space. The missing cone, on the other hand, specifically refers to a block of missing data near the origin of the coordinates in the Fourier space due to the finite NA of objective lens. To minimize the missing-cone artefact, deconvolution using a calculated or measured 3D PSF is typically used. Here we applied a non-negativity constraint in an iterative manner, which replaces negative fluorescence intensities, a numerical artefact, with zero\cite{sung_deterministic_2011}. In the process, new information filling the missing-cone region can be generated as in the Gerchberg-Papoulis algorithm\cite{gerchberg_super-resolution_1974,papoulis_new_1975}. We wrote a code in Matlab (Mathworks, Inc., version 2018a) implementing the inverse projection and iterative application of the non-negativity constraint\cite{kak_principles_1988}. 

\section{Results}
\begin{figure}[htbp]
\centering
\includegraphics[width=\linewidth]{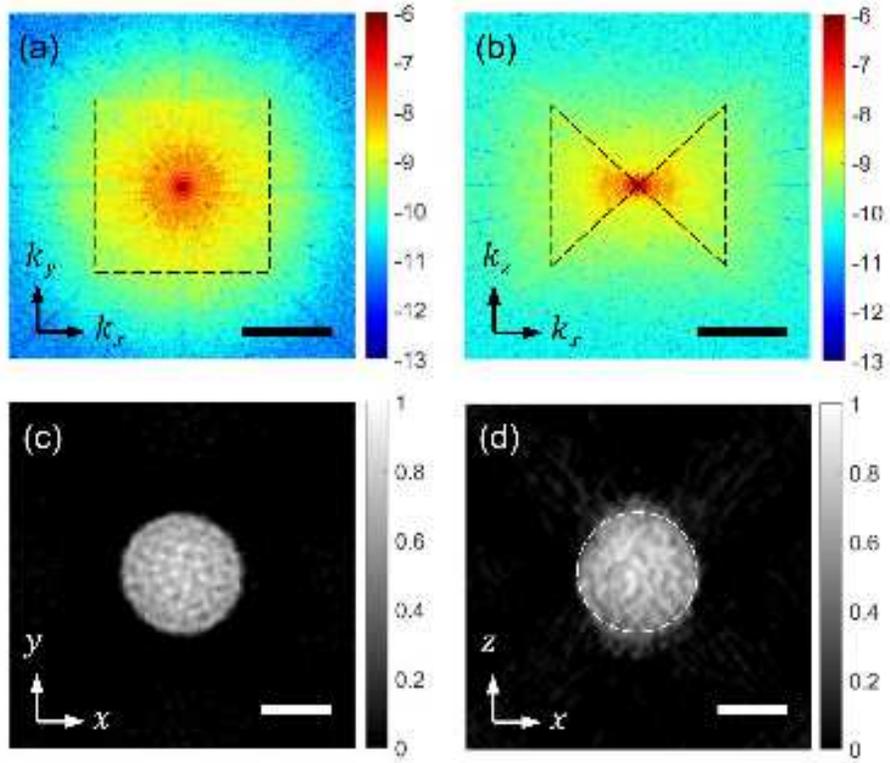}
\caption{Snapshot 3D imaging of a 20-$\mu$m fluorescent bead. (a) and (b) are $k_x-k_y$ and $k_x-k_z$ cross-sections, respectively, of the 3D spatial-frequency spectrum after the data processing. (c) and (d) are the $x-y$ and $x-z$ cross-sections of the 3D fluorescence distribution. Scale bars: 1 $\mu$m$^{-1}$ in (a) and (b) and 10 $\mu$m in (c) and (d)}
\label{fig6}
\end{figure}

Figure~\ref{fig6} shows the horizontal (a) and vertical (b) cross-sections of the 3D spatial-frequency spectrum after completing the inverse projection and 3 iterations to apply the non-negativity constraint. The dotted regions in (a) and (b) represent the data acquired from the measurement, while the regions outside represent the data generated with the deconvolution and the non-negativity constraint. The horizontal (c) and vertical (d) cross-sections of the reconstructed tomogram clearly show the boundary of the spherical bead with a granular pattern inside, which may be attributed to the heterogeneous distribution of fluorochromes. Due to the missing cone, the vertical cross-section in (d) is elongated along the $z$ direction with the intensity spilled out of the dotted circle, which is the true bead boundary estimated from the bead image in (c). This problem pertinent to all wide-field fluorescence imaging techniques may be further alleviated with 3D deconvolution using a measured PSF and stronger constraints such as piecewise smoothness, which is also known as total-variation diminishing algorithm\cite{sung_deterministic_2011}. 

\begin{figure}[htbp]
\centering
\includegraphics[width=\linewidth]{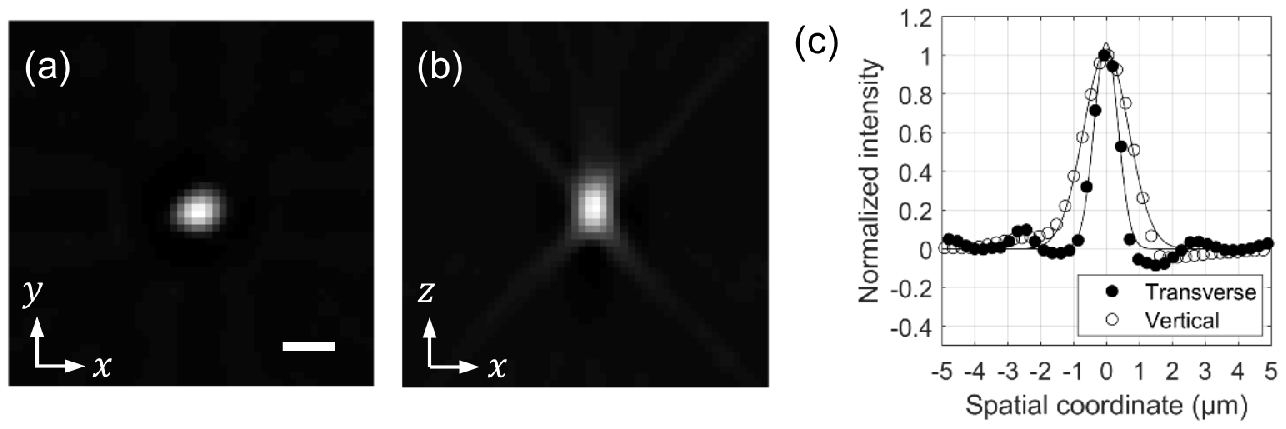}
\caption{Snapshot 3D imaging of a 0.5-$\mu$m fluorescent bead. (a) and (b) are transverse and vertical cross-sections of the reconstructed tomogram. (c) is the intensity profiles along the $x$ and $z$ axes. The scale bar in (a), 2 $\mu$m, also applies to (b).}
\label{fig7}
\end{figure}

Figure~\ref{fig7} shows the horizontal (a) and vertical (b) cross-sections of a 0.5 $\mu$m fluorescent bead, together with the intensity profiles (c) along the $x$ and $z$ axes. The FWHM, which was estimated with a Gaussian fitting, was 0.8 $\mu$m and 1.6 $\mu$m along the $x$ and $z$ directions, respectively. These values, without subtracting the actual bead size, may be considered as a conservative estimate of the transverse and axial resolution of the prototype SPOT system. The measured resolution is slightly higher than the diffraction limit for the effective NA of lenslets, possibly due to the non-negativity constraint and the deconvolution applied to the projection images before the tomographic reconstruction. 

\begin{figure}[htbp]
\centering
\includegraphics[width=\linewidth]{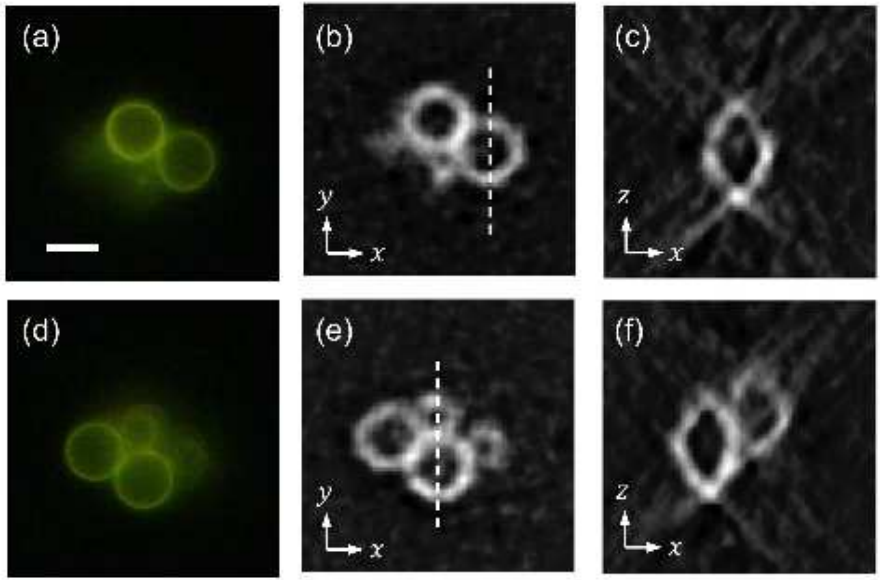}
\caption{Snapshot 3D imaging of a cluster of 6-$\mu$m polystyrene beads with the outermost layer stained with fluorescent dye. (a) and (d): wide-fluorescence images at two axial locations 3 $\mu$m apart. (b) and (e): horizontal cross-sections at the corresponding heights. (c) and (f): vertical cross-sections along the dotted lines in (b) and (e), respectively. The scale bar in (a), 5 $\mu$m, applies to all the other images.}
\label{fig8}
\end{figure}

Figure \ref{fig8} shows wide-field fluorescence images (a and d) of a cluster of 6 $\mu$m fluorescent beads, together with horizontal (b and e) cross-sections of the reconstructed tomogram at the corresponding heights. As can be seen in (a) and (d), there are four beads, two of which are at the same height and the other two are at about 3 $\mu$m below. The horizontal cross-sections obtained with SPOT properly identify the beads at the corresponding axial locations. Applying Gaussian fitting to the measured intensity profile, we measured the thickness of the fluorescent ring to be 1.2 $\mu$m, which corresponds with the diffraction limit, but is slightly thicker than the value measured with a 0.5 $\mu$m bead. This difference may be attributed to the misalignment between the projection images, which can be mitigated using a fiducial marker. The images in (c) and (f) are the vertical cross-sections along the dotted lines in (b) and (e), respectively. They clearly show the ring structure, although the images are elongated along the optical axis, $z$ axis, due to the missing cone. The image in (f) confirms the different axial locations of the beads.

\begin{figure}[htbp]
\centering
\includegraphics[width=\linewidth]{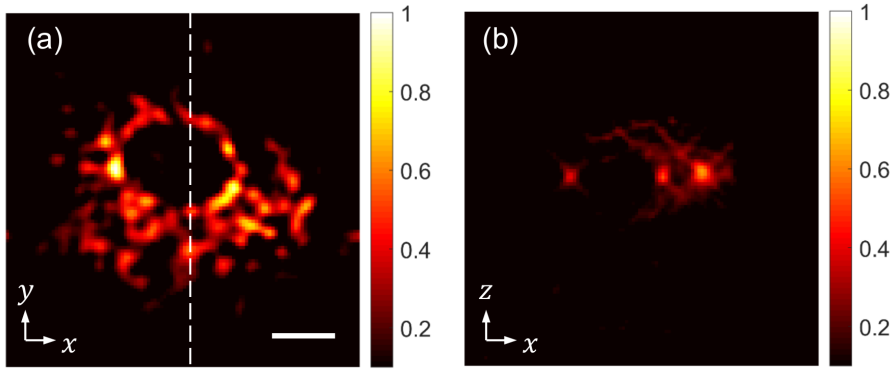}
\caption{Snapshot 3D imaging of mitochondria within a BPAE cell: (a) a horizontal cross-section and (b) a vertical cross-section along the dashed line in (a). The scale bar in (a), 10 $\mu$m, also applies to (b).}
\label{fig9}
\end{figure}

Next, with SPOT, we imaged a bovine pulmonary artery endothelial (BPAE) cell (Invitrogen, F36924) with the mitochondria, F-actin, and DNA stained with red, green, and blue fluorescent dyes, respectively. The triple-band filter cube passed all the three colors, but the red light emitted from the mitochondria was dominantly bright. Figure~\ref{fig9} shows a horizontal (a) and a vertical (b) cross-section of the reconstructed tomogram. The vertical cross-section (b) along the dashed line in (a) shows the mitochondria localized within the thin volume between the microscope slide and the cover slip, which confirms the optical sectioning capability of SPOT. The non-uniform brightness may be attributed to the nonuniform density of mitochondria or different accumulation rate of the fluorophores, which depends on the membrane potential of mitochondria.

\section{Discussion}
The spatial resolution of fluorescence microscopy has dramatically improved and it is now possible to achieve a resolution below 10 nm using several optical techniques\cite{cremer_perspectives_2016}. Despite the development of those super-resolution microscopes and other recent techniques, achieving high spatial and temporal resolution at the same time is still challenging. Several plenoptic configurations have been proposed to address this challenge\cite{llavador_resolution_2016,cong_rapid_2017,scrofani_fimic:_2018}. In this work, we presented a plenoptic configuration for 3D snapshot fluorescence imaging, which we named as snapshot projection optical tomography (SPOT). Using a prototype SPOT system, we imaged different types of fluorescent beads and the mitochondria within a cell. The measurements confirmed the capability of SPOT for imaging specimens with an extended fluorophore distribution as well as isolated fluorochromes. We have also measured the 3D PSF of SPOT using a 0.5 $\mu$m fluorescent bead, which provided the transverse and axial resolutions of 0.8 $\mu$m and 1.6 $\mu$m, respectively. Further, we have developed a forward imaging model for SPOT, based on which an image processing and tomographic reconstruction algorithm was developed. To mitigate the missing-cone problem, we tested the non-negativity constraint, which may be combined with various \textit{a priori} constraints\cite{blumensath_iterative_2009}. 

As with SPOT, FiMic/FLFM and XLFM use MLA as a tube lens, and thus are clearly distinguished from existing LFM, which uses MLA as a Shack-Hartmann sensor. SPOT is distinguished from FiMic/FLFM and XLFM in two major aspects. First, SPOT is dual telecentric, i.e., both object- and image-side telecentric, while FiMic/FLFM and XLFM are only partially object-side telecentric. FiMic/FLFM place an MLA at the Fourier plane (i.e., the back focal plane of objective lens)\cite{llavador_resolution_2016}; however, there is no aperture stop for each lenslet; thus, the system does not have full object-side telecentricity. XLFM has an aperture array installed at the back focal plane; however, two groups of lenslets are axially shifted from the plane\cite{cong_rapid_2017}, which compromises the object-side telecentricity. In contrast, SPOT arranges the MLA with the objective lens in a 4-f configuration and places an iris diaphragm at the back focal plane of a relay lens. The single iris diaphragm serves as the aperture stop for all the lenslets, and it is placed at the plane conjugate to both the back focal plane of objective lens and the front focal plane of lenslets. Thereby, it is dual telecentric. Second, SPOT can block high-angle rays from the fluorophores outside the field of view using the iris diaphragm, which serves as the aperture stop for all the lenslets, without blocking the image-forming rays, as is shown in Fig.~\ref{fig1}. FiMic/FLFM do not have true aperture stops, as the rims of lenslets do not block any rays while the objective aperture is too large. XLFM has an aperture array, which blocks some of the stray rays at the cost of reduced light collection efficiency.
\begin{figure}[htbp]
\centering
\includegraphics[width=\linewidth]{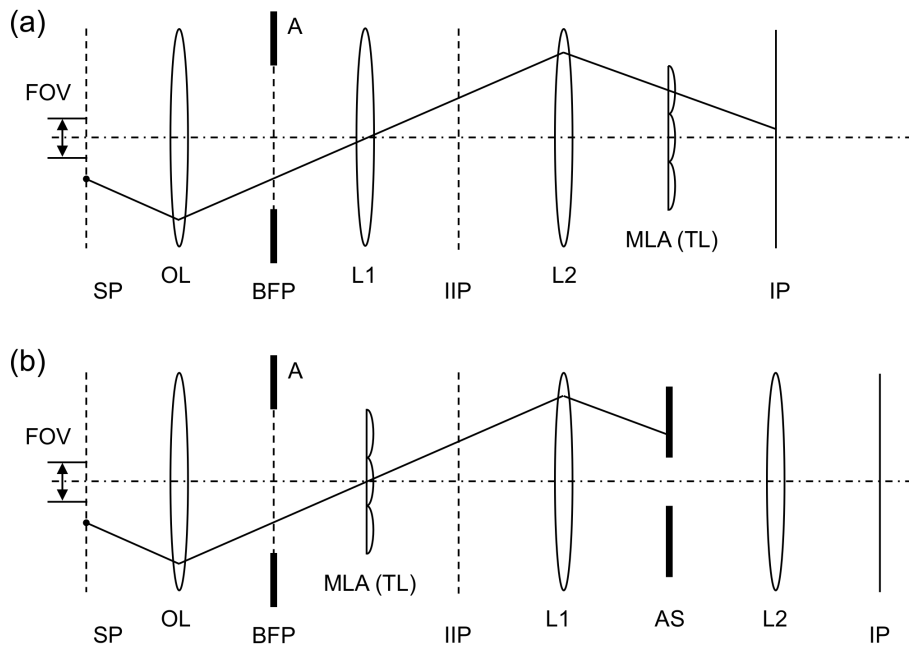}
\caption{Propagation of a high-angle ray in FiMic/FLFM (a) and SPOT (b). An example emission beam path from a fluorophore outside the field of view is shown. SP: sample plane; BFP: back focal plane of the objective lens; IIP: intermediate image plane; IP: image plane; OL: objective lens; A: objective aperture; MLA: microlens array; TL: tube lens; L1 and L2: lenses; FOV: field of view; AS: and aperture stop. The dash-dot line represents the optical axis of objective lens.}
\label{fig10}
\end{figure}

\begin{figure}[htbp]
\centering
\includegraphics[width=\linewidth]{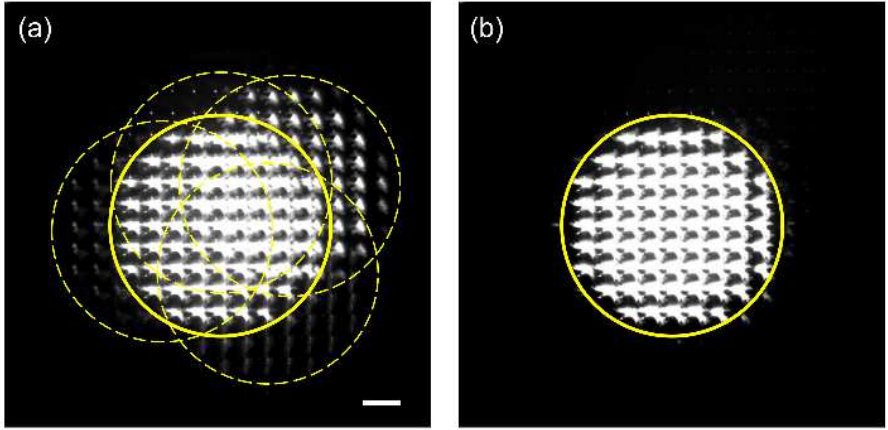}
\caption{Raw SPOT images acquired for different sizes of the aperture stop; the diameter of iris diaphragm was adjusted to 36 mm in (a) and 5 mm in (b). The intensity scale was adjusted to clearly show the ghost images. The scale bar in (a), 100 $\mu$m, also applies to (b).}
\label{fig11}
\end{figure}

Figure~\ref{fig10} compares the trajectories of a high-angle ray in FiMic/FLFM and SPOT, which was emitted from a fluorophore outside the field of view (FOV). For simplicity, unit magnification is assumed. The FOV of both FiMic/FLFM and SPOT is determined by the size of the lenslet as well as its focal length. The objective aperture typically serves as the aperture stop in wide-field imaging. However, in FiMic/FLFM, the angular aperture is divided to record multiple projection images, and the aperture stop required for each projection image is much smaller than the objective aperture. Thus, the objective aperture does not serve as the aperture stop for each lenslet, and the high-angle ray in Fig.~\ref{fig10}(a) can propagate through one of the lenslets and reach the camera sensor. Figure~\ref{fig10}(b) shows the propagation of the same high-angle ray in SPOT, which is blocked by the aperture stop. These high-angle rays from the outside of FOV can create ghost images, which may interfere with the image of the sample within the FOV. Figure~\ref{fig11}(a) shows a raw SPOT image acquired with the iris diaphragm fully opened to the maximum diameter (36 mm). The region marked with the solid line shows the projection images of a single cell in the FOV. The intensity scale has been adjusted to clearly show the ghost images, thereby saturating the pixels in the main projection images. The figure also shows four partially-overlapping groups (marked with dotted lines) of ghost images of neighboring cells. When the diameter of the iris diaphragm is reduced to about 5 mm, SPOT removes the ghost images without affecting the FOV, as can be seen in Fig.~\ref{fig11}(b).

Snapshot 3D imaging techniques including SPOT typically divide the spatial-bandwidth product of the camera to record the sample's depth information\cite{sung_snapshot_2019}. This means a smaller FOV, a lower spatial resolution, or a lower angular resolution. This problem, however, can be alleviated using a high-resolution camera and an iterative reconstruction incorporating \textit{a priori} information of the sample. Using a high-resolution camera, the imaging throughput may be compromised, which is ultimately limited by the camera frame rate or the data transfer speed to computer, whichever is slower. Importantly, using a snapshot 3D technique such as SPOT, the volumetric fluorophore distribution can be recorded instantaneously, which is important for observing a fast, continuous motion in 3D such as that of freely moving \textit{C. elegans}\cite{shaw_three-dimensional_2018}. Although the photons are distributed to the projection images, each projection image records accumulated fluorescence signal. Thus, the signal-to-noise ratio for SPOT can be comparable to that for high-resolution wide-field imaging, when the fluorophore distribution is not sparse. In addition to fluorescence signal, SPOT can be applied to record other luminescence signal such as bioluminescence, electroluminescence, radioluminescence, and phosphorescence\cite{ronda_luminescence:_2007}. 

\section*{Acknowledgments}
This research was funded by the Research Growth Initiative of the University of Wisconsin-Milwaukee, the National Science Foundation (1808331), and the National Institute of General Medical Sciences of the National Institutes of Health (R21GM135848). The content is solely the responsibility of the author and does not necessarily represent the official views of the National Institutes of Health.

\bibliographystyle{IEEEtran}
\bibliography{ms}

\end{document}